\documentstyle[12pt,epsfig]{article}

\def\PRL{\em Phys. Rev. Lett.}

\def\be{\begin{equation}}
\def\ee{\end{equation}}
\def\bea{\begin{eqnarray}}
\def\eea{\end{eqnarray}}

\begin{document}
\vspace*{4cm}
\begin{center}
{\bf STATUS OF THE STANDARD MODEL AND BEYOND}\\
\vspace*{1cm}
{\bf G. Altarelli }\\
\vspace*{0.3cm}
Theory Division, CERN, CH--1211 Geneva 23, Switzerland\\
\vspace*{2cm}
{\bf Abstract}
\end{center}
We present a concise review of the status of the Standard Model and of the search for new physics.

\def\be{\begin{equation}}
\def\ee{\end{equation}}
\def\bea{\begin{eqnarray}}
\def\eea{\end{eqnarray}}
\def\beq{\begin{equation}}
\def\eeq{\end{equation}}
\def\bq{\begin{quote}}
\def\eq{\end{quote}}
\def\gappeq{\mathrel{\rlap {\raise.5ex\hbox{$>$}} {\lower.5ex\hbox{$\sim$}}}}
\def\lappeq{\mathrel{\rlap{\raise.5ex\hbox{$<$}} {\lower.5ex\hbox{$\sim$}}}}
\def\PR{{\it Phys.~Rev.~}}
\def\PRL{{\it Phys.~Rev.~Lett.~}}
\def\NP{{\it Nucl.~Phys.~}}
\def\NPBPS{{\it Nucl.~Phys.~B (Proc.~Suppl.)~}}
\def\PL{{\it Phys.~Lett.~}}
\def\PRep{{\it Phys.~Rep.~}}
\def\AP{{\it Ann.~Phys.~}}
\def\CMP{{\it Comm.~Math.~Phys.~}}
\def\JMP{{\it J.~Math.~Phys.~}}
\def\NC{{\it Nuovo~Cim.~}}
\def\SJNP{{\it Sov.~J.~Nucl.~Phys.~}}
\def\SPJETP{{\it Sov.~Phys.~J.E.T.P.~}}
\def\ZP{{\it Zeit.~Phys.~}}
\def\JP{{\it Jour.~Phys.~}}
\def\JHEP{{\it J.~High~En.~Phys.~}}
\def\vol#1{{\bf #1}}
\def\vyp#1#2#3{\vol{#1} (#2) #3}
\def\Im{\,\hbox{Im}\,}
\def\Re{\,\hbox{Re}\,}
\def\tr{\,{\hbox{tr}}\,}
\def\Tr{\,{\hbox{Tr}}\,}
\def\det{\,{\hbox{det}}\,}
\def\Det{\,{\hbox{Det}}\,}
\def\neath#1#2{\mathrel{\mathop{#1}\limits_{#2}}}
\def\ker{\,{\hbox{ker}}\,}
\def\dim{\,{\hbox{dim}}\,}
\def\ind{\,{\hbox{ind}}\,}
\def\sgn{\,{\hbox{sgn}}\,}
\def\mod{\,{\hbox{mod}}\,}
\def\apm#1{\hbox{$\pm #1$}}
\def\epm#1#2{\hbox{${\lower1pt\hbox{$\scriptstyle +#1$}}
\atop {\raise1pt\hbox{$\scriptstyle -#2$}}$}}
\def\neath#1#2{\mathrel{\mathop{#1}\limits_{#2}}}
\def\gsim{\mathrel{\rlap{\lower4pt\hbox{\hskip1pt$\sim$}}
    \raise1pt\hbox{$>$}}}         
\def\eg{{\it e.g.}}
\def\ie{{\it i.e.}}
\def\viz{{\it viz.}}
\def\etal{{\it et al.}}
\def\rhs{right hand side}
\def\lhs{left hand side}
\def\toinf#1{\mathrel{\mathop{\sim}\limits_{\scriptscriptstyle {#1\rightarrow\infty }}}}
\def\tozero#1{\mathrel{\mathop{\sim}\limits_{\scriptscriptstyle {#1\rightarrow0 }}}}
\def\frac#1#2{{{#1}\over {#2}}}
\def\half{\hbox{${1\over 2}$}}\def\third{\hbox{${1\over 3}$}}
\def\quarter{\hbox{${1\over 4}$}}
\def\smallfrac#1#2{\hbox{${{#1}\over {#2}}$}}
\def\pbp{\bar{\psi }\psi }
\def\vevpbp{\langle 0|\pbp |0\rangle }
\def\as{\alpha_s}
\def\tr{{\rm tr}}\def\Tr{{\rm Tr}}
\def\eV{{\rm eV}}\def\keV{{\rm keV}}
\def\MeV{{\rm MeV}}\def\GeV{{\rm GeV}}\def\TeV{{\rm TeV}}
\def\MS{\hbox{$\overline{\rm MS}$}}
\def\blackbox{\vrule height7pt width5pt depth2pt}
\def\matele#1#2#3{\langle {#1} \vert {#2} \vert {#3} \rangle }
\def\VertL{\Vert_{\Lambda}}\def\VertR{\Vert_{\Lambda_R}}
\def\Real{\Re e}\def\Imag{\Im m}
\def\bp{\bar{p}}\def\bq{\bar{q}}\def\br{\bar{r}}
\catcode`@=11 
\def\slash#1{\mathord{\mathpalette\c@ncel#1}}
 \def\c@ncel#1#2{\ooalign{$\hfil#1\mkern1mu/\hfil$\crcr$#1#2$}}
\def\lsim{\mathrel{\mathpalette\@versim<}}
\def\gsim{\mathrel{\mathpalette\@versim>}}
 \def\@versim#1#2{\lower0.2ex\vbox{\baselineskip\z@skip\lineskip\z@skip
       \lineskiplimit\z@\ialign{$\m@th#1\hfil##$\crcr#2\crcr\sim\crcr}}}
\catcode`@=12 
\def\twiddles#1{\mathrel{\mathop{\sim}\limits_
                        {\scriptscriptstyle {#1\rightarrow \infty }}}}

\section{Precision Tests of the Standard Model} The results of the electroweak precision tests as well as of the searches
for the Higgs boson and for new particles performed at LEP and SLC are now available in nearly final form.  Taken together
with the measurements of $m_t$, $m_W$ and the searches for new physics at the Tevatron, and with some other data from low
energy experiments, they form a very stringent set of precise constraints to compare with the Standard Model (SM) or with
any of its conceivable extensions \cite{LEPEW}. When confronted with these results, on the whole the SM performs rather
well, so that it is fair to say that no clear indication for new physics emerges from the data. There are, however, a few
results that show a 2-3$\sigma$ deviation from the SM fit: for example, the NuTeV anomaly \cite{NuT}, atomic parity
violation (APV) in Cs
\cite{APV} and the muon g-2 \cite{BNL}. 

The NuTev experiment is a precise measurement of $s^2_W=1-m_W^2/m_Z^2$ from the
Pascos-Wolfenstein ratio of neutral to charged current deep inelastic cross-sections from $\nu_{\mu}$ and
$\bar{\nu}_{\mu}$ using the TeVatron beams. Allegedly the result shows a 3$\sigma$ deviation from the SM prediction.  In
my opinion the NuTeV anomaly could simply arise from an underestimation of the theoretical error in the QCD analysis
needed to extract $s^2_W$. The related cuisine is not very transparent and there are a number of critical points. In fact the
QCD lowest order (LO) parton formalism on which the analysis was based is too crude to match the experimental accuracy (it was
sufficient in previous experiments of the same type, but is not adequate now that the experimental errors are so much
smaller). In particular a small asymmetry in the momentum carried by the strange and antistrange quarks, $s-\bar s$, could
have a large effect
\cite{DFGRS}. The collaboration claims to have measured this quantity from dimuons. But a LO analysis of $s-\bar s$ is
inadequate and the result cannot be directly trasplanted here. In fact neglected non leading corrections of order
$\alpha_s$*valence are large and process dependent. A tiny violation of isospin symmetry in parton distributions, too small to
be seen elsewhere, can similarly be of some importance. 

For APV the quantity $Q_W$ which is usually "measured" and compared with the SM
prediction is actually an idealised pseudo-observable that corresponds for the naive value for N neutrons and Z protons
in the nucleus. For Cs the theoretical best fit value \cite{LEPEW} is $(Q_W)_{th}=-72.880\pm 0.003$. The "experimental"
value actually contains a variety of QED and nuclear effects that keep changing all the time. And indeed since the last
fit \cite{LEPEW} (showing a ~1.5$\sigma$ deviation) a new evaluation of the QED corrections led to
$(Q_W)_{th}=-72.71\pm 0.49$ \cite{KF}. So in this very moment APV is OK!. 

If we remove NuTeV and APV from the fit the
resulting $\chi^2$ is reasonably good but not perfect $\chi^2/dof=18.2/13$. The ~3$\sigma$ deviation shown by the BNL'02
data on 
$(g-2)_{\mu}$ \cite{BNL}
is not included in the list of precision tests that correspond to that $\chi^2$ value. In units $10^{-10}$ for
$(g-2)_{\mu}$ we have (exp-th)$=30.4\pm 11.1$ or (exp-th)$=36.1\pm 10.9$ depending on the evaluation of hadronic
contributions from exclusive or inclusive cross-section data \cite{HMNT}, respectively. In fact, the main uncertainties on
the prediction arise from the hadronic corrections to the photon vacuum polarization and to light-by-light scattering. To
give an idea, the contribution of LO hadronic corrections to vacuum polarization is (in the same units) $683.1\pm 6.2$ (from inclusive data)
while the light-by light hadronic contribution is $8\pm 4$ (this is the term that changed of sign last year when a
mistake was corrected: it was $-8.5\pm 2.5$). I see no obvious objection to this result and it is well possible that it could
indeed arise from some new physics effect.

\begin{figure}
\begin{center}
\epsfig{file=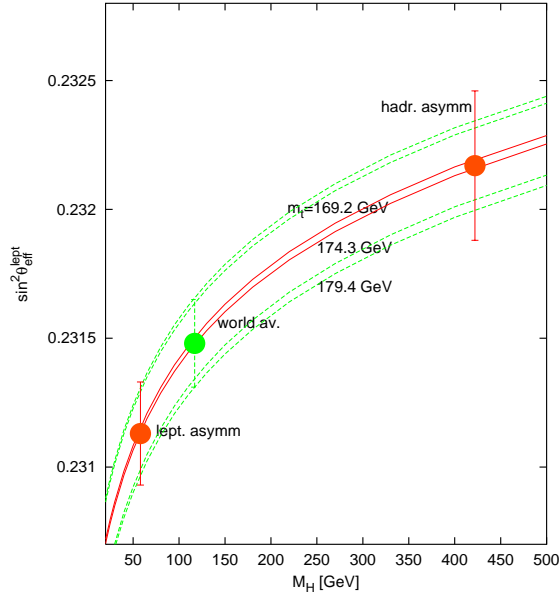,width=0.55\textwidth}
\vskip1.0cm
\caption{\label{fig0}
The data for $\sin^2\theta_{\rm eff}$ are plotted vs $m_H$. The values of $(\sin^2\theta_{\rm eff})_l$,
$(\sin^2\theta_{\rm eff})_h$ and their combination (as if they could be combined) are shown each at the
$m_H$ value that would corresponds to it given the central value of $m_t$ (I owe this figure to P. Gambino).}
\end{center}
\end{figure}

Even leaving aside NuTeV, APV and $(g-2)_{\mu}$, if we look
at the results in detail, there are a number of features that are either not satisfactory or could indicate the presence
of small new physics effects. 
One problem is that the two most precise measurements of
$\sin^2\theta_{\rm eff}$ from $A_{LR}$ and $A^b_{FB}$ differ by
$\sim 3\sigma$. More in general, there appears to be a discrepancy between $\sin^2\theta_{\rm eff}$ measured from leptonic
asymmetries ($(\sin^2\theta_{\rm eff})_l$) and from hadronic asymmetries ($(\sin^2\theta_{\rm eff})_h$). In fact the result
from
$A_{LR}$ is in good agreement with the leptonic asymmetries measured at LEP, while all hadronic asymmetries are better
compatible with the result of
$A^b_{FB}$. The situation is shown in Fig.~\ref{fig0}.
It is well known that this discrepancy is not easily explained in terms of some new physics effect in the
$b
\bar b Z$ vertex. In fact $A^b_{FB}$ is the product of lepton- and $b$-asymmetry factors:
$A^b_{FB}\propto A_\ell A_b$, where
$A_f=(g^{f2}_L-g^{f2}_R)/(g^{f2}_L+g^{f2}_R)$ with $g_{R,L}=g_V\pm g_A$.  The sensitivity of $A^b_{FB}$ to $A_b$ is
limited, because the
$A_\ell$ factor is small, so that  a rather large change of the
$b$ couplings with respect to the SM is needed in order to reproduce the measured discrepancy (precisely a $\sim 30\%$ change
in the right-handed coupling $g_R$, an effect too large to be a loop effect but which could be produced at the tree level,
e.g. by mixing of the b quark with a new heavy vectorlike quark \cite{CTW}). But then this effect should affect the direct
measurement of
$A_b$ performed at SLD using the LR polarized $b$ asymmetry, even within the moderate precision of this result, and it
should also appear in the accurate measurement of $R_b\propto {g^{b}_R}^2+{g^{b}_L}^2$. Neither $A_b$ nor $R_b$ confirm the
need of a new effect. One concludes that most probably the observed discrepancy is due to a large statistical fluctuation or an
experimental problem. In any case this discrepancy implies that the ambiguity in the measured value of
$\sin^2\theta_{\rm eff}$ is larger than the nominal error obtained from averaging all the existing determinations.

Another point of focus is the relation between the fitted Higgs mass and the lower limit on this mass from direct
searches, $m_H > 114$~GeV \cite{cha}, \cite{ACGGR}.  In particular, given the experimental value of $m_t$, the measured
result for $m_W$ (with perfect agreement between LEP and the Tevatron) is somewhat large. The values of $m_W$ and
$\sin^2\theta_{\rm eff}$ measured from leptonic asymmetries, taken together with the results on the
$Z_0$ partial widths, push the central value of $m_H$ very much down. In fact, if one arbitrarily excludes
$(\sin^2\theta_{\rm eff})_h$, the fitted value of $m_H$ would only become marginally
consistent, to a level that depends on the adopted value and the error for
$\alpha_{QED}(m_Z)$. Consistency is reinstaured if the results from hadronic asymmetries are also included, because they
drive the fitted $m_H$ value towards somewhat larger values \footnote{In a recent update of the LEP analysis 
\cite{moriond} the central value of $m_W$ went down, and the world average moved from $m_W=80.449(34)$ down to $m_W=80.426(34)$
so that the problem is less acute.  Note that if the central value of $m_t$ increases, the problem is alleviated.}. These statements can be made more quantitative by restricting our attention to only the
most important observables:
$m_t$,
$m_W$,
$\Gamma_l$,
$R_b$, ,
$\sin^2\theta_{\rm eff}$, $\alpha_s(m_Z)$,
$\alpha_{QED}(m_Z)$. Then, for $\alpha^{-1}_{QED}(m_Z)= 128.936\pm 0.049$, one obtains the following results \cite{ACGGR}.
Taking both $(\sin^2\theta_{\rm eff})_l$and $(\sin^2\theta_{\rm eff})_h$ as separate inputs one obtains $\chi^2/dof=18.4/4$,
$m_{Hcentral}\sim 100~GeV$ and $m_H\lappeq 212~GeV$ at $95\%$. If we only keep $(\sin^2\theta_{\rm eff})_h$ we have
$\chi^2/dof=15.3/3$: not much advantage because of the friction between $(\sin^2\theta_{\rm eff})_h$ and $m_W$. If we only
keep $(\sin^2\theta_{\rm eff})_l$ then we obtain $\chi^2/dof=2.5/3$, $m_{Hcentral}\sim 42~GeV$ and $m_H\lappeq 109~GeV$ at
$95\%$: the $\chi^2$ is good but the clash with the direct limit on $m_H$ emerges. But recall that in all these fits
one assumes the validity of the SM. However we have good reasons (summarized later in this talk) to believe that new
physics must be present near the weak scale. While the effects from physics beyond the SM are certainly not prominent,
given that we cannot clearly identify them, still we must keep in mind that these effects could distort the analysis with
contributions to some observables of the order of the present errors. In particular the discrepancy between
$(\sin^2\theta_{\rm eff})_l$ and $(\sin^2\theta_{\rm eff})_h$ could be due to some modification of the $Z\rightarrow b \bar
b$ vertex, although this appears difficult for the reasons explained above. Or if the $\sin^2\theta_{\rm eff}$ problem is
due to a fluctuation or an experimental problem, then some small new physics effect could reconcile the precision tests
with the limits from direct searches of the Higgs particle. We will now explore this second possibility in the context of
supersymmetric (SUSY) extensions of the SM, in particular in the Minimal Supersymmetric Standard Model (MSSM). While
there are no relevant modifications of the $Z\rightarrow b \bar b$ vertex in this class of models,
the quality of the fit and the consistency with the direct limit on
$m_H$ can be improved with respect to the SM, even in the most unfavourable case for the SM that the results on
$\sin^2\theta_{\rm eff}$ from the hadronic asymmetries are discarded. We have found \cite{ACGGR} that if sleptons and to a
lesser extent charginos and neutralinos, have masses close to their present experimental limits it is possible to
considerably improve the overall picture. In particular the possible MSSM effects are substantially increased if we allow
the sneutrino masses to be as small as allowed by the direct limits on
$m^2_{\tilde{\nu}}$ and by those on charged slepton masses, which are related by
$m^2_{\tilde{\ell}^-}=m^2_{\tilde\nu}+m^2_W|\cos 2\beta|$. At moderately large values of $\tan\beta$ (\ie\ for
$|\cos 2\beta|\sim 1$), light sneutrinos with masses as low as
$50-70$~GeV are not excluded by present limits, while charged sleptons must be heavier than about
$100$~GeV. These low values of the sneutrino mass can still be compatible with the neutralino being the lightest
supersymmetric particle. We recall that $\tan\beta\gappeq 2-3$ is required by LEP, and large
$\tan\beta$ and light sleptons are indicated by the possible deviation observed by the recent Brookhaven
result~\cite{BNL} on the muon $g-2$, if this discrepancy is to be explained by a MSSM effect. 

For this analysis in the MSSM we use the technique of the epsilon parameters $\epsilon_1$,
$\epsilon_2$, $\epsilon_3$ and $\epsilon_b$~\cite{epsilon}.  For "oblique" contributions from new
physics (that is terms arising from vacuum polarization diagrams) the variations of $\epsilon_1$, $\epsilon_2$ and
$\epsilon_3$ are proportional to the shifts of the
$T$, $U$, and $S$ parameters, respectively \cite{stu}. But for the MSSM not all important contributions are oblique. We
recall that the starting point of the epsilon analysis is a one-to-one definition of the
$\epsilon_i$ in terms of four basic observables that were chosen to be $\sin^2\theta_{\rm eff}$ from
$A^{\mu}_{FB}$,
$\Gamma_{\mu}$, $m_W$ and $R_b$. Given the experimental values of these quantities, the corresponding experimental values
of the
$\epsilon_i$ follow, independent of $m_t$ and $m_H$, with an error that also includes the effect of the present ambiguities in
$\alpha_s(m_Z)$ and
$\alpha_{QED}(m_Z)$. If one assumes lepton universality, which is well supported by the data within the present accuracy,
then the combined results on $\sin^2\theta_{\rm eff}$ from all leptonic asymmetries can be adopted together with the
combined leptonic partial width
$\Gamma_\ell$. At this level the epsilon analysis is model independent within the stated lepton universality assumption.
As a further step we can observe that by including the information on the hadronic widths arising from
$\Gamma_Z$, $\sigma_h$, $R_\ell$ the central values of the
$\epsilon_i$ are not much changed (with respect to the error size) but the errors are decreased.  Different is the case
of including the results from the hadronic asymmetries in the combined value of
$\sin^2\theta_{\rm
  eff}$. In this case, obviously, the determination of $\epsilon_i$ is sizeably affected and one remains with the
alternative between an experimental problem or a bizarre effect of some new physics (not present in the MSSM). But if we
remain within the first stage of purely leptonic measurements plus $m_W$ ($R_b$ and other related observables like $A_b$
essentially only affect
$\epsilon_b$, which is secondary for the present purposes and will be disregarded in the following), the
$\epsilon_i$ analysis is quite general and, in particular, is independent of an assumption of oblique correction
dominance. 

\begin{figure}[htb]
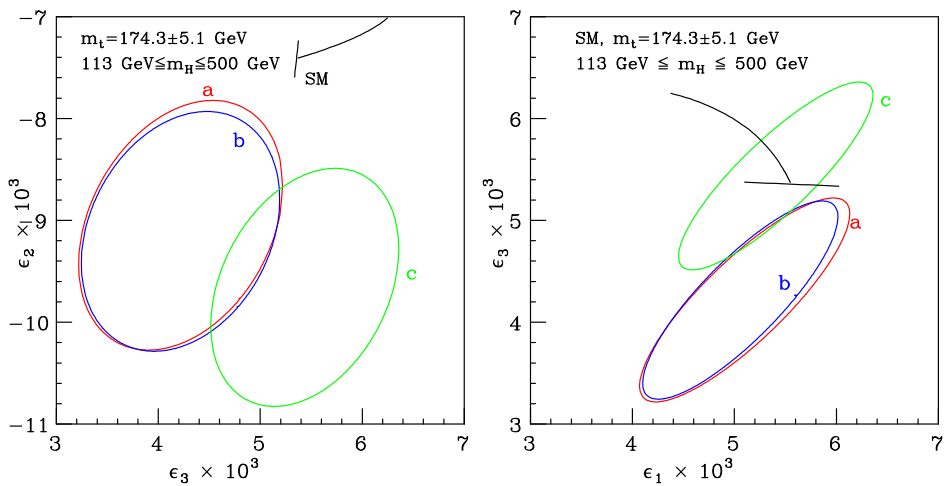

\begin{center}
\epsfig{file=fig1-l.ps,width=0.45\textwidth}
\epsfig{file=fig1-r.ps,width=0.45\textwidth}
\caption{\label{fig1}
One-sigma ellipses in the $\epsilon_3-\epsilon_2$ (left)
and in the $\epsilon_1-\epsilon_3$ (right) planes obtained from:
{\rm a}.~$m_W$, $\Gamma_\ell$, $\sin^2\theta_{\rm eff}$ from all leptonic
asymmetries, and $R_b$; {\rm b}.~the same observables, plus the hadronic 
partial widths derived from $\Gamma_Z$, $\sigma_h$ and $R_\ell$; 
{\rm c}.~as in {\rm b}.,
but with $\sin^2\theta_{\rm eff}$ also including the hadronic
asymmetry results. The solid straight lines represent
the SM predictions for $m_H=113$~{\rm GeV}
and $m_t$ in the range $174.3\pm 5.1$~{\rm GeV}. The dotted curves represent
the SM predictions for $m_t=174.3$~{\rm GeV} and $m_H$ in the range $113$ to 
$500$~{\rm GeV}.}
\end{center}
\end{figure}

The comparison with the SM can be repeated in the context of the
$\epsilon_i$. The predicted theoretical values of the $\epsilon_i$ in the SM depend on $m_H$ and
$m_t$, while are practically independent of
$\alpha_s(m_Z)$ and $\alpha_{QED}(m_Z)$. For $m_H=114$~GeV and
$m_t=174$~GeV the results are shown in Fig.~\ref{fig1}. One finds that the experimental value of $\epsilon_1$ agrees within
errors with the SM prediction. The agreement between fitted value and prediction for
$\epsilon_1$, which, contrary to $\epsilon_2$ and $\epsilon_3$, contains a quadratic dependence on
$m_t$, reflects the fact that the fitted value of $m_t$ is in perfect agreement with the measured value. 
The other variable that depends quadratically on
$m_t$ is
$\epsilon_b$ (not shown in the figures). The agreement of the fitted and predicted values of
$\epsilon_b$ reflects the corresponding present normality of the results for $R_b$.The
experimental value of 
$\epsilon_2$ is below the theoretical expectation by about
$1~\sigma$. We recall that $m_W$ is related to $\epsilon_2$ and the fact that the experimental value is below
the prediction for this quantity corresponds to the statement that
$m_W$ would prefer a value of $m_H$ much smaller than $m_H=114$~GeV. The value of $\epsilon_3$ is in agreement with the SM
only if $\sin^2\theta_{\rm
  eff}$ is taken as the average of the leptonic and hadronic measurements (because only in this case we have full
compatibility with the limit on $m_H$). If instead we only take $(\sin^2\theta_{\rm eff})_l$ as an input then
$\epsilon_3$ is below the prediction for $m_H=114$~GeV by about $1~\sigma$. In summary, we reobtain in the language of
the $\epsilon_i$ the same picture than from the direct SM fit.

\begin{figure}[h]
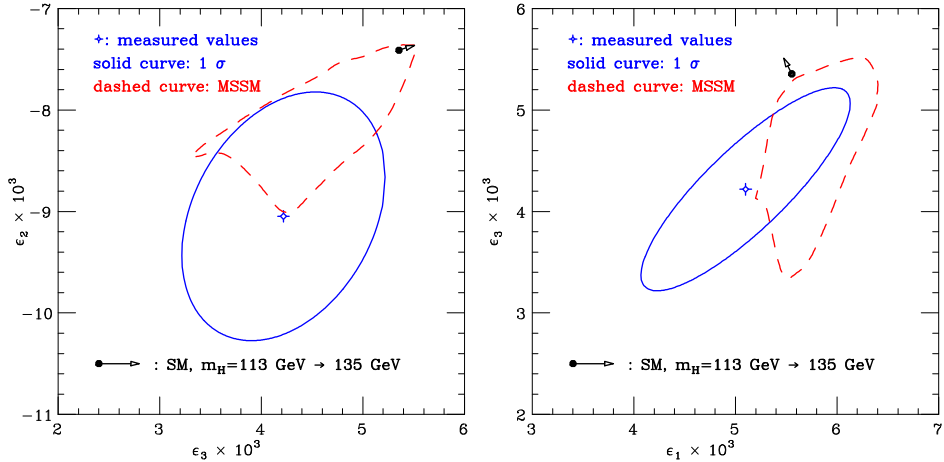

\begin{center}
\epsfig{file=fig2-l.ps,width=0.45\textwidth}
\epsfig{file=fig2-r.ps,width=0.45\textwidth}
\caption{\label{fig2}\
Measured values (cross) of $\epsilon_3$ and $\epsilon_2$ (left)
and of $\epsilon_1$ and $\epsilon_3$ (right), with their
$1~\sigma$ region (solid ellipses), corresponding to
case {\rm a} of fig.~\ref{fig1}. The area inside the
dashed curves represents the MSSM prediction for $m_{{\tilde e}_L}$ between
96 and 300~{\rm GeV}, $m_{\chi^+}$ between 105 and 300~{\rm GeV},
$-1000$~{\rm GeV}~$<\mu<1000$~{\rm GeV}, $\tan\beta=10$, 
$m_{{\tilde e}_L}=1$~{\rm TeV}. and $m_A=1$~{\rm TeV}.}
\end{center}
\end{figure}

Now we want to investigate whether low-energy supersymmetry can reconcile a Higgs mass above the direct experimental
limit with a good
$\chi^2$ fit of the electroweak data, even in the case of
$\sin^2\theta_{\rm eff}$ near the value obtained from leptonic asymmetries. Our approach is to discard the measurement of
$A^b_{FB}$, which cannot be reproduced by conventional new physics effects, fix the Higgs mass above its present limit,
and look for supersymmetric corrections that can fake a very light SM Higgs boson. As seen from Fig.~\ref{fig2}, this can
be achieved if the new physics contributions to the $\epsilon$ parameters amount to shifting $\epsilon_2$ and $\epsilon_3$
down by slightly more than $1~\sigma$, while leaving
$\epsilon_1$ essentially unchanged.
Squark loops cannot induce this kind of shifts in the $\epsilon$ parameters, since their leading effect is a positive
contribution to
$\epsilon_1$. Thus, we assume that all squarks are heavy, with masses of the order of one TeV. Since the mass of the
lightest Higgs
$m_H$ receives a significant contribution from stop loops, we can treat $m_H$ as an independent parameter and, in our
analysis, we fix
$m_H=113$~GeV. Varying the pseudoscalar Higgs mass $m_A$ does not modify the results of our fit, and therefore we fix
$m_A=1$~TeV. The choice of the right-handed slepton mass has also an insignificant effect on the fit. Therefore, we are
left with four relevant supersymmetric free parameters: the weak gaugino mass $M_2$, the higgsino mass $\mu$, the ratio
of the Higgs vacuum expectation values $\tan\beta$ (which are needed to describe the chargino--neutralino sector), and a
supersymmetry-breaking mass for the left-handed sleptons, ${\tilde m}_{\ell_L}$ (lepton flavour universality is assumed).
The choice of the $B$-ino mass parameter $M_1$ does not significantly affect our results and, for simplicity, we have
assumed the gaugino unification relation $M_1=
\frac{5}{3} M_2 \tan^2\theta_W $. Figure~\ref{fig2} shows the range of the $\epsilon$ parameters that can be spanned by varying
$M_2$, $\mu$, $\tan\beta$, and ${\tilde m}_{\ell_L}$, consistently with the present experimental constraints. We have
imposed a limit on charged slepton masses of 96~GeV~, on chargino masses of 103~GeV~, and on the cross section for
neutralino production
$\sigma(e^+e^-\to\chi_1^0\chi_2^0\to\mu^+\mu^-\slash{E})<0.1$~pb. We have also required that the supersymmetric
contribution to the muon anomalous magnetic moment, $a_\mu=(g-2)/2$, lie within the range
$0<\delta a_\mu <7.5\times 10^{-9}$.  As apparent from Fig.~\ref{fig2}, light particles in the chargino--neutralino sector and
light left-handed sleptons shift the values of $\epsilon_i$ in the favoured direction, and by a sufficient amount to
obtain a satisfactory fit. As for the mass spectrum responsible for this effects on the $\epsilon$ parameters, the most
significant contribution is coming from light sneutrinos. As already mentioned, the effect is maximal when $\tan\beta$ is large
since this allows the smallest possible sneutrino mass compatible with the charged slepton mass bound,
$m_{{\tilde e}_L}^2=m_{\tilde\nu}^2+m_W^2\left|\cos 2\beta\right|$.
Values as low as $m_{\tilde\nu}\sim 50-80~GeV$ are in general allowed by present limits, and lead to the largest shifts
of $\epsilon_i$.

In conclusion, we have presented an example of how new physics effects could affect the interpretation of electroweak
precision tests and perhaps explain some of the features that we observe.

\section{Outlook on Avenues beyond the Standard Model}

Given the success of the SM why are we not satisfied with that theory? Why not just find the Higgs
particle, for completeness, and declare that particle physics is closed? The reason is that there are
both conceptual problems and phenomenological indications for physics beyond the SM. On the conceptual side the most
obvious problems are that quantum gravity is not included in the SM and the related hierarchy problem. Among the main
phenomenological hints for new physics we can list coupling unification, dark matter, neutrino masses, 
baryogenesis and the cosmological vacuum energy.

The computed evolution with energy
of the effective SM gauge couplings clearly points towards the unification of the electro-weak and strong forces (Grand
Unified Theories: GUT's) at scales of energy
$M_{GUT}\sim  10^{15}-10^{16}~ GeV$ which are close to the scale of quantum gravity, $M_{Pl}\sim 10^{19}~ GeV$.  One is led to
imagine  a unified theory of all interactions also including gravity (at present superstrings provide the best attempt at such
a theory). Thus GUT's and the realm of quantum gravity set a very distant energy horizon that modern particle theory cannot
ignore. Can the SM without new physics be valid up to such large energies? This appears unlikely because the structure of the
SM could not naturally explain the relative smallness of the weak scale of mass, set by the Higgs mechanism at $\mu\sim
1/\sqrt{G_F}\sim  250~ GeV$  with $G_F$ being the Fermi coupling constant. This so-called hierarchy problem is related to
the
presence of fundamental scalar fields in the theory with quadratic mass divergences and no protective extra symmetry at
$\mu=0$. For fermion masses, first, the divergences are logarithmic and, second, they are forbidden by the $SU(2)\bigotimes
U(1)$ gauge symmetry plus the fact that at
$m=0$ an additional symmetry, i.e. chiral  symmetry, is restored. Here, when talking of divergences, we are not
worried of actual infinities. The theory is renormalisable and finite once the dependence on the cut off is
absorbed in a redefinition of masses and couplings. Rather the hierarchy problem is one of naturalness. Given that the SM cannot be the fundamental theory, we should see the
cut off as a parameterization of our ignorance on the new physics that will modify the theory at large energy
scales. Then it is relevant to look at the dependence of physical quantities on the cut off and to demand that no
unexplained enormously accurate cancellations arise. 

The hierarchy problem ca be put in very practical terms: loop corrections to the higgs mass squared are
quadratic in $\Lambda$. The most pressing problem is from the top loop. With $m_h^2=m^2_{bare}+\delta m_h^2$ the top loop
gives 
\beq
\delta m_{h|top}^2\sim \frac{3G_F}{\sqrt{2} \pi^2} m_t^2 \Lambda^2\sim (0.3\Lambda)^2 \label{top}\\ 
\eeq
If we demand that the correction does not exceed the light Higgs mass indicated by the precision tests, $\Lambda$ must be
close, $\Lambda\sim o(1~TeV)$. Similar constraints arise from the quadratic $\Lambda$ dependence of loops with gauge bosons and
scalars, which, however, lead to less pressing bounds. So the hierarchy problem demands new physics to be very close (in
particular the mechanism that quenches the top loop). Actually, this new physics must be rather special, because it must be
very close, yet its effects are not clearly visible (the "LEP Paradox" \cite{BS}). Examples of proposed classes of solutions
for the hierarchy problem are:

¥ $\bf{Supersymmetry.}$ In the limit of exact boson-fermion symmetry the quadratic divergences of bosons must cancel so that
only log divergences remain. However, exact SUSY is clearly unrealistic. For approximate SUSY (with soft breaking terms),
which is the basis for all practical models, $\Lambda$ is replaced by the splitting of SUSY multiplets, $\Lambda\sim
m_{SUSY}-m_{ord}$. In particular, the top loop is quenched by partial cancellation with s-top exchange. 

¥ $\bf{Technicolor.}$ The Higgs system is a condensate of new fermions. There are no fundamental scalar Higgs sector, hence no
quadratic divergences associated to the $\mu^2$ mass in the scalar potential. This mechanism needs a very strong binding force,
$\Lambda_{TC}\sim 10^3~\Lambda_{QCD}$. It is  difficult to arrange that such nearby strong force is not showing up in
precision tests. Hence this class of models has been disfavoured by LEP, although pockets of resistance still exist (for
recent reviews, see, for example, \cite{L-C}). 

¥ $\bf{Large~compactified~extra~dimensions.}$ The idea is that $M_{Pl}$ appears very large, that is gravity seems very weak, 
because we are fooled by hidden extra dimensions so that the real gravity scale is reduced down to
$o(1~TeV)$. This possibility is very exciting in itself and it is really remarkable that it is compatible with experiment.

¥ $\bf{"Little~Higgs" models.}$ In these models extra symmetries allow $m_h\not= 0$ only at two-loop level, so that $\Lambda$
can be as large as
$o(10~TeV)$ with the Higgs within present bounds (the top loop is quenched by exchange of heavy vectorlike new charge-2/3
quarks).

We now briefly comment in turn on these possibilities.

SUSY models are the most developed and most widely accepted. Many theorists consider SUSY as established at the Planck
scale $M_{Pl}$. So why not to use it also at low energy to fix the hierarchy problem, if at all possible? It is interesting
that viable models exist. The necessary SUSY breaking can be introduced through soft
terms that do not spoil the good convergence properties of the theory. Precisely those terms arise from
supergravity when it is spontaneoulsly broken in a hidden sector. This is the case of the MSSM \cite{Martin}. The MSSM 
is a completely specified,
consistent and computable theory which is compatible with all precision electroweak tests. In this
most traditional approach SUSY is broken in a hidden sector and the scale of SUSY breaking is very
large of order
$\Lambda\sim\sqrt{G^{-1/2}_F M_{Pl}}$. But since the hidden sector only communicates with the visible sector
through gravitational interactions the splitting of the SUSY multiplets is much smaller, in the TeV
energy domain, and the Goldstino is practically decoupled. 
But alternative mechanisms of SUSY breaking are also being considered. In one alternative scenario \cite{gau} the (not so
much) hidden sector is connected to the visible one by ordinary gauge interactions. As these are much
stronger than the gravitational interactions, $\Lambda$ can be much smaller, as low as 10-100
TeV. It follows that the Goldstino is very light in these models (with mass of order or below 1 eV
typically) and is the lightest, stable SUSY particle, but its couplings are observably large. The radiative
decay of the lightest neutralino into the Goldstino leads to detectable photons. The signature of photons comes
out naturally in this SUSY breaking pattern: with respect to the MSSM, in the gauge mediated model there are typically
more photons and less missing energy. The main appeal of gauge mediated models is a better protection against
flavour changing neutral currents but naturality problems tend to increase. As another possibility it has been
pointed out that there are pure gravity contributions to soft masses that arise from gravity theory
anomalies \cite{ano}. In the assumption that these terms are dominant the associated spectrum and phenomenology have been
studied. In this case gaugino masses are proportional to gauge coupling beta functions, so that the gluino is much heavier
than the electroweak gauginos, and the wino is most often the lightest SUSY particle. 

What is really unique to SUSY with respect to all other extensions of the SM listed above is that the MSSM or
similar models are well defined and computable up to $M_{Pl}$ and, moreover, are not only compatible but actually
quantitatively supported by coupling unification and GUT's. At present the most direct
phenomenological evidence in favour of supersymmetry is obtained from the unification of couplings in GUTs.
Precise LEP data on $\alpha_s(m_Z)$ and $\sin^2{\theta_W}$ show that
standard one-scale GUTs fail in predicting $\sin^2{\theta_W}$ given
$\alpha_s(m_Z)$ and $\alpha(m_Z)$ while SUSY GUTs are in agreement with the present, very precise,
experimental results. If one starts from the known values of
$\sin^2{\theta_W}$ and $\alpha(m_Z)$, one finds \cite{LP} for $\alpha_s(m_Z)$ the results:
$\alpha_s(m_Z) = 0.073\pm 0.002$ for Standard GUTs and $\alpha_s(m_Z) = 0.129\pm0.010$ for SUSY-GUTs
to be compared with the world average experimental value $\alpha_s(m_Z) =0.119\pm0.003$. Another great asset of SUSY GUT's
is that proton decay is much slowed down with respect to the non SUSY case. First, the unification mass $M_{GUT}\sim~{\rm few}
10^{16}~GeV$, in typical SUSY GUT's, is about 20-30 times larger than for ordinary GUT's. This makes p decay via gauge
boson exchange negligible and the main decay amplitude arises from dim-5 operators with higgsino exchange, leading to a
rate close to but still compatible with existing bounds (see, for example,\cite{AFM}).

In spite of these virtues it is true that the lack of SUSY signals at LEP and the lower limit on $m_H$ pose problems
for the MSSM. The lightest Higgs particle is predicted in the MSSM to be below $m_h\lappeq~130~GeV$. The limit on the SM
Higgs
$m_H\gappeq~114~GeV$ considerably restricts the available parameter space of the MSSM requiring relatively large $\tan\beta$
($\tan\beta\gappeq~2-3$: at tree level $m^2_h=m^2_Z\cos^2{2\beta}$) and rather heavy s-top (the loop corrections
increase with $\log{\tilde{m_t^2}}$). Stringent naturality constraints also follow from imposing that the electroweak
symmetry breaking occurs at the right place: in SUSY model the breaking is induced by the running of the $H_u$ mass
starting from a common scalar mass $m_0$ at $M_{GUT}$. The squared Z mass $m_Z^2$ can be expressed as a linear
combination of the SUSY parameters $m_0^2$, $m_{1/2}^2$, $A^2_t$, $\mu^2$,... with known coefficients. Barring
cancellations that need fine tuning, the SUSY parameters, hence the SUSY s-partners, cannot be too heavy. The LEP limits,
in particular the chargino lower bound $m_{\chi+}\gappeq~100~GeV$, are sufficient to eliminate an important region of the
parameter space, depending on the amount of allowed fine tuning. For example, models based on gaugino universality at the
GUT scale are discarded unless a fine tuning by at least a factor of ~20 is not allowed. Without gaugino
universality \cite{kane} the strongest limit remains on the gluino mass: $m_Z^2\sim 0.7~m_{gluino}^2+\dots$ which is still
compatible with the present limit $m_{gluino}\gappeq~200~GeV$.

The non discovery of SUSY at LEP has given further impulse to the quest for new ideas in beyond the SM. Large extra
dimensions \cite{Jo} and "little Higgs" \cite{schm} models are the most interesting new directions in model building. Large
extra dimension models propose to solve the hierarchy problem by bringing gravity down from $M_{Pl}$ to $m\sim~o(1~TeV)$ where
$m$ is the string scale. Inspired by string theory one assumes that some compactified extra dimensions are sufficiently large
and that the SM fields are confined to a 4-dimensional brane immersed in a d-dimensional bulk while gravity, which feels the
whole geometry, propagates in the bulk. We know that the Planck mass is large because gravity is weak: in fact $G_N\sim
1/M_{Pl}^2$, where
$G_N$ is Newton constant. The idea is that gravity appears so weak because a lot of lines of force escape in extra
dimensions. Assume you have $n=d-4$ extra dimensions with compactification radius $R$. For large distances, $r>>R$, the
ordinary Newton law applies for gravity: $F\sim G_N/r^2\sim 1/(M_{Pl}^2r^2)$. At short distances,
$r\lappeq R$, the flow of lines of force in extra dimensions modifies Gauss law and $F^{-1}\sim m^2(mr)^{d-4}r^2$. By
matching the two formulas at $r=R$ one obtains $(M_{Pl}/m)^2=(Rm)^{d-4}$. For $m\sim~1~TeV$ and $n=d-4$ one finds that
$n=1$ is excluded ($R\sim 10^{15} cm$), for $n=2~R$  is at the edge of present bounds $R\sim~1~ mm$, while for $n=4,6$,
$R\sim~10^{-9}, 10^{-12}~cm$.  In all these models a generic feature is the occurrence of Kaluza-Klein (KK) modes.
Compactified dimensions with periodic boundary conditions, as for quantization in a box, imply a discrete spectrum with
momentum $p=n/R$ and mass squared $m^2=n^2/R^2$. There are many versions of these models. The SM brane can itself have a
thickness $r$ with $r<\sim~10^{-17}~cm$ or $1/r>\sim~1~TeV$, because we know that quarks and leptons are pointlike down to
these distances, while for gravity there is no experimental counter-evidence down to $R<\sim~0.1~mm$ or
$1/R>\sim~10^{-3}~eV$. In case of a thickness for the SM brane there would be KK recurrences for SM fields, like $W_n$,
$Z_n$ and so on in the $TeV$ region and above. In any case there are the towers of KK recurrences of the graviton. They
are gravitationally coupled but there are a lot of them that sizably couple, so that the net result is a modification
of cross-sections and the presence of missing energy. There are models with factorized metric ($ds^2=\eta_{\mu
\nu}dx^{\mu}dx^{\nu}+h_{ij}(y)dy^idy^j$, where y (i,j) denotes the extra dimension coordinates (and indices), or models
with warped metric ($ds^2=e-{2kR|\phi|} \eta_{\mu \nu}dx^{\mu}dx^{\nu}-R^2\phi^2$ \cite{RS}.

Large extra dimensions provide a very exciting scenario. Already it is remarkable that this possibility is
compatible with experiment. However, there are a number of criticisms that can be brought up. First, the hierarchy problem is
more translated in new terms rather than solved. In fact the basic relation $Rm=(M_{Pl}/m)^{2/n}$ shows that $Rm$, which one
would apriori expect to be $0(1)$, is instead ad hoc related to the large ratio $M_{Pl}/m$. The suppression by the possibly
small exponent $2/n$ could  be illusory because the right quantity could be the volume and not the radius. Also it is
not clear how extra dimensions can by themselves solve the LEP paradox (the large top loop corrections should be
controlled by the opening of the new dimensions and the onset of gravity): since
$m_H$ is light
$\Lambda\sim 1/R$ must be relatively close. But precision tests put very strong limits on $\Lambda$ In fact in typical
models of this class there is no mechanism to sufficiently quench the corrections. Perhaps some additional ingredient must be added. No simple, realistic model has
yet emerged as a benchmark. But it is attractive to imagine that large extra dimensions could be a part of the truth,
perhaps coupled with some additional symmetry or even SUSY.

Let us make a short digression on the cosmological constant or vacuum energy problem \cite{tu}. The exciting recent results
on cosmological parameters, culminating with the precise WMAP measurements \cite{WMAP}, have shown that vacuum energy accounts
for about 2/3 of the critical density: $\Omega_{\Lambda}\sim 0.65$, Translated into familiar units this means for the energy
density $\rho_{\Lambda}\sim (2~10^{-3}~eV)^4$ or $(0.1~mm)^{-4}$. It is really interesting (and not at all understood)
that $\rho_{\Lambda}^{1/4}\sim \Lambda_{EW}^2/M_{Pl}$ (close to the range of neutrino masses). It is well known that in
field theory we expect $\rho_{\Lambda}\sim \Lambda_{cutoff}^4$. If the cut off is set at $M_{Pl}$ or even at $0(1~TeV)$
there would an enormous mismatch. In exact SUSY $\rho_{\Lambda}=0$, but SUSY is broken and in presence of breaking 
$\rho_{\Lambda}^{1/4}$ is in general not smaller than the typical SUSY multiplet splitting. Another closely related
problem is "why now?": the time evolution of the matter or radiation density is quite rapid, while the density for a
cosmological constant term would be flat. If so, them how comes that precisely now the two density sources are
comparable. This suggests that the vacuum energy is not a cosmological constant term, buth rather the vacuum expectation
value of some field (quintessence) and that the "why now?" problem is solved by some dynamical mechanism.

In "little Higgs" models the symmetry of the SM is extended to a suitable global group G that also contains some
gauge enlargement of $SU(2)\bigotimes U(1)$, for example $G\supset [SU(2)\bigotimes U(1)]^2\supset SU(2)\bigotimes
U(1)$. The Higgs particle is a pseudo-Goldstone boson of G that only takes mass at 2-loop level, because two distinct
symmetries must be simultaneously broken for it to take mass,  which requires the action of two different couplings in
the same diagram. Then in the relation
between
$\delta m_h^2$ and
$\Lambda^2$ there is an additional coupling and an additional loop factor that allow for a bigger separation between the Higgs
mass and the cut-off. Typically, in these models one has one or more Higgs doublets at $m_h\sim~0.2~TeV$, and a cut-off at
$\Lambda\sim~10~TeV$. The top loop quadratic cut-off dependence is partially canceled, in a natural way guaranteed by the
symmetries of the model, by a new coloured, charge-2/3, vectorial quark $\chi$ of mass around $1~TeV$ (a fermion not a scalar
like the s-top of SUSY models). Certainly these models involve a remarkable level of group theoretic virtuosity. However, in
the simplest versions sofar proposed one is faced with problems with precision tests of the SM \cite{prob}. Even with
vectorlike new fermions large corrections to the epsilon parameters arise from exchanges of the new gauge bosons
$W'$ and $Z'$ (due to lack of custodial $SU(2)$ symmetry). In order to comply with these constraints the cut-off must be
pushed towards large energy and the amount of fine tuning needed to keep the Higgs light is still quite large.
Probably these bad features can be fixed by some suitable complication of the model. But, in my opinion, the real limit of
this approach is that it only offers a postponement of the main problem by a few TeV, paid by a complete loss of
predictivity at higher energies. In particular all connections to GUT's are lost. 

\section{Summary and Conclusion}

Supersymmetry remains the standard way beyond the SM. What is unique to SUSY, beyond leading to a set of consistent and
completely formulated models, as, for example, the MSSM, is that this theory can potentially work up to the GUT energy scale.
In this respect it is the most ambitious model because it describes a computable framework that could be valid all the way
up to the vicinity of the Planck mass. The SUSY models are perfectly compatible with GUT's and are actually quantitatively
supported by coupling unification and also by what we have recently learned on neutrino masses. All other main ideas for going
beyond the SM do not share this synthesis with GUT's. The SUSY way is testable, for example at the LHC, and the issue
of its validity will be decided by experiment. It is true that we could have expected the first signals of SUSY already at
LEP, based on naturality arguments applied to the most minimal models (for example, those with gaugino universality at
asymptotic scales). The absence of signals has stimulated the development of new ideas like those of large extra dimensions
and "little Higgs" models. These ideas are very interesting and provide an important referfence for the preparation of LHC
experiments. Models along these new ideas are not so completely formulated and studied as for SUSY and no well defined and
realistic baseline has sofar emerged. But it is well possible that they might represent at least a part of the truth and it
is very important to continue the exploration of new ways beyond the SM.

\vfill
\end{document}